\documentclass[prl,showpacs,superscriptaddress,twocolumn]{revtex4}
\usepackage{graphicx,amssymb,amsmath,psfrag,color}

\begin{document}

\newcommand{\secref}[1]{\aref{sec:#1}.~fejezet}
\newcommand{\figref}[1]{\aref{fig:#1}.~\'abra}
\newcommand{\figreff}[1]{\aref{fig:#1}.~\'abr\'a}
\newcommand{\tabref}[1]{\aref{tab:#1}.~t\'abl\'azat}
\newcommand{\egyref}[1]{\aref{eq:#1}.~egyenlet}
\newcommand{\Secref}[1]{\Aref{sec:#1}.~fejezet}
\newcommand{\Figref}[1]{\Aref{fig:#1}.~\'abra}
\newcommand{\Figreff}[1]{\Aref{fig:#1}.~\'abr\'a}
\newcommand{\Tabref}[1]{\Aref{tab:#1}.~t\'abl\'azat}
\newcommand{\Egyref}[1]{\Aref{eq:#1}.~egyenlet}
\newcommand{\kv}[0]{\mathbf{k}}
\newcommand{\Rv}[0]{\mathbf{R}}
\newcommand{\rv}[0]{\mathbf{r}}
\newcommand{\al}[0]{\mathbf{a_{1}}}
\newcommand{\as}[0]{\mathbf{a_{2}}}
\newcommand{\K}[0]{\mathbf{K}}
\newcommand{\Kp}[0]{\mathbf{K'}}
\newcommand{\dkv}[0]{\delta\kv}
\newcommand{\dkx}[0]{\delta k_{x}}
\newcommand{\dky}[0]{\delta k_{y}}
\newcommand{\dk}[0]{\delta k}
\newcommand{\cv}[0]{\mathbf{c}}
\newcommand{\qv}[0]{\mathbf{q}}
\newcommand{\Rr}[0]{\Rv_{\rv}}

\title{Mean-field quantum phase transition in graphene and in general gapless systems}
\author{\'Ad\'am B\'acsi}
\affiliation{Department of Physics, Budapest University of Technology and
  Economics, Budafoki \'ut 8, 1111 Budapest, Hungary}
\author{Attila Virosztek}
\affiliation{Department of Physics, Budapest University of Technology and
  Economics, Budafoki \'ut 8, 1111 Budapest, Hungary}
\affiliation{Research Institute for Solid State Physics and Optics, PO Box 49, 1525 Budapest, Hungary}
\author{L\'aszl\'o Borda}
\affiliation{Department of Theoretical Physics, Budapest University of Technology and
  Economics, Budafoki \'ut 8, 1111 Budapest, Hungary}
\author{Bal\'azs D\'ora}
\email{dora@kapica.phy.bme.hu}
\affiliation{Department of Physics, Budapest University of Technology and
  Economics, Budafoki \'ut 8, 1111 Budapest, Hungary}

\date{\today}

\begin{abstract}
We study the quantum critical properties of antiferromagnetism in graphene at $T=0$ within mean-field (MF) theory. The resulting exponents differ from the conventional MF exponents, describing finite temperature
transitions.  Motivated  by this, we have developed the MF theory of general gapless phases with density of states $\rho(\varepsilon)\sim |\varepsilon|^r$, $r>-1$, with the interaction as control parameter.
For $r>2$, the conventional MF exponents \`a la Landau are recovered, while for $-1<r<2$, the exponents  vary significantly with $r$. 
The critical interaction is finite for $r>0$, therefore  no weak-coupling solution exists in this range.
This generalizes the 
results on quantum criticality of the gapless Kondo systems to bulk correlated phases.
\end{abstract}


\maketitle

Graphene, a single sheet of carbon atoms, arranged in a hexagonal lattice provided us with lots of excitement recently\cite{castro07}.
Its novel electronic properties arise from the linear energy-momentum dispersion of electrons at low energies, forming a Dirac cone, resulting in
a
linearly vanishing density of states (DOS) around half filling, similarly to d-wave superconductors. It explains the semi-metallic response of graphene, 
placing it somewhere between good metals with a constant DOS and insulators with a clean gap.
Although most of its electronic and transport properties can be understood by analyzing the Dirac equation under appropriate conditions,
the effect of electron-electron interaction on its Dirac quasiparticles remains a bit mysterious\cite{netopauling}. 
No conclusive evidence for any kind of phase transition of either pristine or doped graphene has been reported to date,
and in spite of the increasing amount of theoretical work\cite{marteloGW,gusynin,paivaQPT,drutlahdeMC3,feldnerMF,drutlahdeMC,yazyev,peresmagnetic,drutlahdeMC2,liliu,guineaRG}, no consensus is reached on the role of interactions.

In addition to graphene, other systems with semi-metallic behaviour are also available.
In the graphene family, trilayer graphene with appropriate stacking exhibits a diverging DOS\cite{multilayer,MM08} as $\sim |\varepsilon|^{-1/3}$.
Around a two dimensional semi-Dirac point, such as the one in VO$_2$/TiO$_2$ nano-heterostructures\cite{pickett}, the DOS varies as $\sim \sqrt{|\varepsilon|}$.
Disorder belonging to specific classes can modify the exponent of the DOS of Dirac electrons to almost arbitrary power law\cite{morita} as $\sim |\varepsilon|^r$ with $0<r<1$. 
These can also be generalized to three dimensions\cite{volovik}, providing us with a whole zoo of systems exhibiting a power-law DOS around the Fermi energy.

Systems with gapless DOS ($\rho(\varepsilon)\sim |\varepsilon|^r$ with $r>-1$) are expected to modify the behaviour of correlated phases and alter their critical behaviour.
While quantum impurity models (i.e. Kondo model) in metallic hosts with a power law DOS have been extensively investigated\cite{bullarmp} following the pioneering work of 
Withoff and Fradkin\cite{WF}, much less is known about the interplay of bulk correlations (e.g. Hubbard model) and power law DOS except for
isolated efforts on specific models. 
On general grounds, the power-law vanishing DOS implies restricted phase space for quasiparticles, which should weaken the effect of interactions, 
analogously to low-dimensional systems (i.e. phase transitions are less likely).

This motivates us to study the critical properties of bulk correlated systems with a power law DOS within the mean field (MF) theory.
Our interest is twofold: on the one hand, we want to understand how and under which conditions correlations can alter
dramatically the physical properties of graphene (e.g. by opening a gap in the spectrum) at $T=0$;
on the other hand, the quantum critical properties of general bulk gapless phases have never been systematically studied (except for the Kondo problem, which is local) 
and our work
intends to fill this gap.

We start with the Hamiltonian of graphene in tight-binding approximation, written as
\begin{equation}
H_{0}=\sum_{\kv\sigma}\left[\begin{array}{c c} a^{+}_{\kv\sigma} & b^{+}_{\kv\sigma}\end{array}\right]\left[\begin{array}{c c} 0 & tf^{*}(\kv)\\ tf(\kv) & 0\end{array}\right]\left[\begin{array}{c} 
a_{\kv\sigma}\\ b_{\kv\sigma} \end{array}\right]
\end{equation}
where $t$ is the hopping integral and $f(\kv)=1+e^{i\kv\al}+e^{i\kv\as}$ with the lattice vectors $\al$ and $\as$, $a_{\kv\sigma}$ and $b_{\kv\sigma}$ are 
the annihilation operators of the Bloch states 
corresponding to the $A$ and $B$ sublattices, respectively\cite{castro07}. 
The electron spectrum vanishes linearly at the Dirac points as $\varepsilon_{0}(\kv)=\pm\hbar v_{F} k$ at the corners of the Brillouin zone ($\K$, $\Kp$), 
$v_{F}$ is the Fermi velocity. 
The DOS is linear in energy $\rho(\varepsilon)=2|\varepsilon|/(\pi\hbar^{2}v_{F}^{2})$. 
Since it differs significantly from that in normal metals (i.e. $\rho(\varepsilon)\sim$const),
it is natural to ask how the  interplay of interactions and the above linear DOS can modify the critical dynamics, which serves as the basic motivation for our study.
For the sake of simplicity, we focus on antiferromagnetism within MF approximation in the presence of a simple repulsive on-site Hubbard interaction, so that 
$s=\langle\sum_{\Rv}\hat{s}(\Rv)\rangle/N$ is the 
expectation value of the staggered magnetization where $\hat{s}(\Rv)=\sum_\sigma \sigma(a^{+}_{\Rv\sigma}a_{\Rv\sigma}-b^{+}_{\Rv\sigma}b_{\Rv\sigma})/4$ 
and $N$ the number of unit 
cells. 
The MF theory for the Hubbard model gives
\begin{gather}
H_{MF}=\sum_{\kv\sigma}\left[\begin{array}{c c} a^{+}_{\kv\sigma} & b^{+}_{\kv\sigma}\end{array}\right]\left[\begin{array}{c c} -Us\sigma & tf^{*}(\kv)\\ tf(\kv) & Us\sigma\end{array}\right]
\left[\begin{array}{c} a_{\kv\sigma}\\ b_{\kv\sigma} \end{array}\right]+\nonumber\\
+\frac{UN}{2}+2UNs^{2}
\end{gather}
 which describes quasi-particles with the dispersion relation $\varepsilon(\kv)=\pm\sqrt{\varepsilon^{2}_{0}(\kv)+\Delta^{2}}$, $\Delta=Us$ is the energy gap in the spectrum.
We introduce a cut-off energy $W$, preserving the total number of states as $W=\sqrt{\sqrt{3}\pi}|t|$.

The ground state energy of the half-filled system (pristine graphene) at $T=0$ reads as
\begin{gather}
\frac{E(s)}{N}=\frac{U(1+4s^2)}{2}-\frac{4U^{3}}{3W^{2}}\left[\left(s^{2}+\left(\frac{W}{U}\right)^{2}\right)^{\frac{3}{2}}-|s|^{3}\right],
\label{gsenergy}
\end{gather}
subject to minimization with respect to $s$. As opposed to standard MF results, there is no non-trivial weak coupling solution,
the normal phase with $s=0$ extends until $U<W$. For bigger values of $U$, we enter into the antiferromagnetic phase with 
$|s|=(1-W^2/U^2)/2$, 
the critical value of U being $U_{c}=W\approx 2.33|t|$. 
The accurate MF critical value can be calculated \cite{sorella} using the exact energy spectrum without linearization, yielding
$U_{c}=2.23|t|$. 
The critical exponent of the order parameter $s$ is $\beta=1$ which differs from the usual (thermal) MF value of $1/2$. 
The difference can be understood from the expansion of the total energy for small $s$ where the second order 
term is followed by a third order term (instead of $\sim s^{4}$ term which is usual in Landau theory):
\begin{equation}
\label{eq:landaugraphene}
\frac{E(s)}{N}=\frac{E(0)}{N}-as^{2}+c|s|^{3},
\end{equation} 
with $E(0)=\frac{UN}{2}\left(1-\frac{8W}{3U}\right)$, $a=2U\left(\frac{U}{W}-1\right)$ and $c=\frac{4U^{3}}{3W^{2}}$.
The unconventional third order term is induced by the linear behaviour of the DOS at the Fermi level. 
Near the critical point, 
\begin{equation}
E_{min}(U)=E(0)-2N\left(U-U_{c}\right)^{3}/(3W^2),
\end{equation}
and the critical exponent of the ground state energy is $\alpha=-1$. The control parameter in this quantum phase transition 
is $U$, as opposed to the reduced temperature in classical phase transitions.
In an external magnetic field, we have
$E(s)/N=E(0)/N-as^{2}+c|s|^{3}-hs$ close to the critical point, where $h$ is the staggered field, coupled to the order parameter. 
At $U=U_{c}$ ($a=0$), the extremal value of the energy is at $|s|=(h/(4W))^{1/2}$ so the critical 
exponent of the conjugate field is $\delta=2$.
Note, that for a normal Landau theory $\alpha=0$ and $\delta=3$.

The bare susceptibility of the staggered magnetization at zero temperature is obtained as
$\chi=\frac{1}{2U}\frac{W^{2}}{W^{2}+U^{2}}$,
dressed within RPA as
\begin{equation}
\chi_{RPA}=\frac{1}{2U}\frac{W^{2}}{U^{2}-W^{2}},
\end{equation}
diverging at the critical point, and the critical exponent of the susceptibility is $\gamma=1$. The 
susceptibility can 
be determined as the inverse of the second derivative of the total energy (\ref{eq:landaugraphene}) with respect to $s$ evaluated at the minimum as well. 
This calculation provides $\chi=(2a)^{-1}=W(4U)^{-1}(U-W)^{-1}$ and also yields $\gamma=1$.
The correlation length and the relaxation time can be written as $\xi=\hbar v_{F}/\Delta$ and $\tau=\hbar/\Delta$ so the corresponding exponents are $\nu=1$ and $z=1$.
 The critical exponents calculated above fulfill scaling relations. 
The renormalization group eigenvalues are $y_{t}=1$ and $y_{h}=2$ with the effective dimension $d_{eff}=d+z=3$. The above exponents are summarized in Table \ref{tab:exponents}.

Our results agree qualitatively with the RG analysis in the presence of long-range Coulomb interaction\cite{herbutrelmottcrit,herbutint}, 
namely that around the Dirac point, weak interactions cannot trigger phase transitions.
The critical properties of the above MF theory for antiferromagnetism in graphene agrees with the BCS theory for graphene\cite{uchoa}, whenever comparison was possible.

\begin{table}[h]
\centering
\begin{tabular}{|c|c|c|}
\hline
exponent & definition &Quantum  MF value\\
\hline
$\alpha$ & $E_{min}(U)\sim\left(U-U_{c}\right)^{2-\alpha}$ & $-1$\\
$\beta$ & $|s|\sim\left(U-U_{c}\right)^{\beta}$ & $1$\\
$\gamma$ & $\chi\sim\left(U-U_{c}\right)^{-\gamma}$ & $1$\\
$\delta$ & $|s|\sim h^{1/\delta},\textmd{ }U=U_{c}$ & $2$\\
$\nu$ & $\xi\sim\left(U-U_{c}\right)^{-\nu}$ & $1$\\
$z$ & $\tau\sim\left(U-U_{c}\right)^{-\nu z}$ & $1$\\
\hline
\end{tabular}
\caption{Quantum MF exponents in graphene at $T=0$.}
\label{tab:exponents}
\end{table}

Having determined the MF quantum critical properties of graphene, we observe that the exponents differ from those of a finite temperature thermal MF phase transition.
To generalize these results, let us consider an electron system in $d$ spatial dimension with energy spectrum consisting of two bands, touching each other at the 
Fermi level with DOS
\begin{equation}
\rho(\varepsilon)=C|\varepsilon|^{r}
\end{equation}
where $r>-1$ must hold so that the total number of electrons $N$ does not diverge. We introduce a cut-off energy $W$ so that $N=CW^{r+1}/(r+1)$ holds, and the case 
$r=1$ corresponds to graphene.
Such a DOS follows from a non-interacting spectrum, $\varepsilon_0({\bf k})\sim \pm k^z$, connecting $r$ with the dynamical critical exponent ($z$) as $z(r+1)=d$

By switching on an electron-electron interaction, an ordered phase can develop. Within the MF approximation, an energy gap opens in the quasi-particle dispersion relation 
$\varepsilon({\bf  k})=\pm\sqrt{\varepsilon_0({\bf k})^{2}+\Delta^{2}}$. The control 
parameter of the transition is $U$ and the order parameter can be denoted by $s$ (describing arbitrary ordering with  spin or charge or pairing character),
 which is proportional to the energy gap $\Delta=Us$.
At zero temperature the total energy can usually be written using the DOS and the energy spectrum as
\begin{gather}
\label{eq:totenergy}
\frac{E(\Delta)}{N}=\frac{\Delta^{2}}{U}-\int_{0}^{W}d\varepsilon \sqrt{\varepsilon^{2}+\Delta^{2}}\frac{\rho(\varepsilon)}{N}=\nonumber\\
=\frac{\Delta^2}{U}-\Delta\,\, \textmd{}_2F_1\left(-\frac 
12,\frac{r+1}{2},\frac{r+3}{2};-\frac{W^2}{\Delta^2}\right),
\end{gather}
where $N\Delta^2/U$ is the general energy cost of the ordered phase, $_2F_1$ is the Gauss hypergeometric function. By expanding it with respect to $\Delta/W$ in order to 
determine the critical behaviour, we get 
\begin{equation}
\label{eq:landau0}
\frac{E(\Delta)}{N}=-a(U)\Delta^{2}+c\Delta^{r+2}+b\Delta^{4}
\end{equation} 
measured from the  ground state energy in the normal (unordered) phase, $E(0)=-NW(r+1)/(r+2)$ and the coefficients are 
\begin{gather}
a(U)=\frac{r+1}{2rW}-\frac{1}{U},\\ 
c=\frac{r+1}{W^{r+1}}\frac{\Gamma\left(-\frac{2+r}{2}\right)\Gamma\left(\frac{1+r}{2}\right)}{4\sqrt{\pi}}, \textmd{ }b=\frac{r+1}{8W^{3}(r-2)}
\end{gather}
where $\Gamma$ is the gamma function. Higher order terms in the expansion are higher even powers of $\Delta/W$ ($6,8,...$), not influencing the critical behaviour.
The ground state energy for $r=1$ agrees with Eq. \eqref{eq:landaugraphene} for graphene apart from spin degeneracy.
 The expression of the energy, Eq. \eqref{eq:landau0}, differs from the usual Landau 
expansion because of the term $\Delta^{r+2}$, producing  different critical exponents when $-1<r<2$. 

For $r\geq 2$, we get the usual MF critical exponents ($a(U)$ changes sign and $b>0$ holds in this case),
and the critical coupling $U_c=2rW/(r+1)$. The conventional Landau theory, with the temperature replaced by the interaction, operates.
 For $-1<r<2$, 
the term $\Delta^{4}$ can be neglected in (\ref{eq:landau0}) because close to the critical point it is much smaller than the $\Delta^{2}$ and $\Delta^{r+2}$ terms.

For $0<r<2$, the 
 critical $U$ is obtained from $a(U_{c})=0$ which yields again $U_{c}=2rW/(r+1)$. 
Therefore, as long as $r>0$, there is always a finite critical interaction, and no conventional weak-coupling solution exists.
Due to this, near the critical point,  $a(U)\approx U_{c}^{-2}(U-U_{c})$ so $a(U)$ can be regarded as the deviation of the control parameter from its critical 
value, and its scaling can be used to determine quantum criticality. The total energy is minimal at
\begin{equation}
\Delta=\left(\frac{2a}{c(r+2)}\right)^{1/r},
\end{equation}
giving $\beta=1/r$ for $a>0$ and $\Delta=0$ for $a<0$. In the ordered phase the ground state total energy reads as
\begin{gather}
\frac{E_{min}(U)}{N}=-\frac{r}{r+2}\left(\frac{2}{c(r+2)}\right)^{\frac{2}{r}}a^{\frac{2}{r}+1},
\end{gather}
giving $\alpha=1-\frac 2r$.
Upon introducing an appropriate external field $h$ which is conjugate to the order parameter, we obtain to linear order in small fields
$\Delta=h/(2U|a|)$ in the normal phase ($a<0$) and $\delta\Delta=h/(2Uar)$ in the ordered phase ($a>0$). In both cases,
\begin{gather} 
\chi\sim a^{-1},
\end{gather} 
therefore $\gamma=1$.
At the critical point $a=0$, the total energy is extremal at
\begin{equation}
\Delta=\left(\frac{h}{cU(r+2)}\right)^{1/(r+1)}\sim h^{1/(r+1)},
\end{equation}
yielding $\delta=r+1$. The exponents calculated above fulfill scaling relations and yield the renormalization group 
eigenvalues
$y_{h}=d_{eff}\frac{r+1}{r+2}$, $y_{t}=d_{eff}\frac{r}{r+2}$ where $d_{eff}=d+z=d(r+2)/(r+1)$ is the effective dimension of the system.

This was the regime, often investigated in the pseudogap Kondo problem\cite{bullarmp}. 
However, the $-1<r<0$ region is also possible physically\cite{vojtabulla}, and we focus on its critical behaviour in the followings: $c<0$ and 
the 
coefficient 
$a(U)$ is 
also 
negative for all $U$. The critical point is at $U_{c}=0$ so we have to calculate with the original order parameter $s$, and not $\Delta$. Near the critical 
point the total energy 
reaches its minimum at
\begin{equation}
s=\left(\frac{2}{|c|(r+2)}\right)^{1/r}U^{-1-1/r}
\end{equation}
leading to $\beta=-1-\frac 1r$.
The distinction between $\Delta$ and $s$ is important when $U_c=0$, since $\Delta\sim U^{1/|r|}$, and its identification as the order parameter would lead to incorrect scaling.
The minimum energy scales as
\begin{equation}
\frac{E_{min}(U)}{N}=\frac{r}{r+2}\left(\frac{2}{|c|(r+2)}\right)^{2/r}U^{-1-2/r},
\end{equation}
determining $\alpha=3+\frac 2r$.
The linear response to an external field is $\delta s=-h/(2rU)$ and the susceptibility
diverges with exponent $\gamma=1$, while a finite external field modifies the order parameter as 
$s\sim h^{1+r}$, giving $\delta=1/(1+r)$.

\begin{figure}[!h]
\centering
\psfrag{exponent}[t][b][1][0]{exponents}
\psfrag{r}[b][t][1][0]{$r$}
\includegraphics[width=8cm]{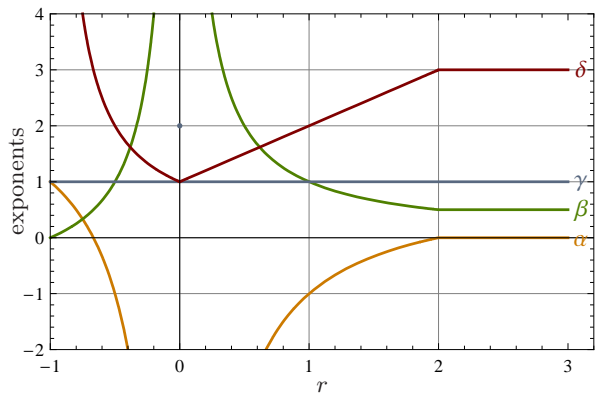}
\caption{The critical exponents as a function of the DOS exponent, $r$ are plotted. For $r>2$, the conventional MF exponents are recovered.}
\label{fig:exps}
\end{figure}

The scaling relations are fulfilled by these exponents
and the renormalization group eigenvalues are
$y_{t}=d_{eff}\frac{|r|}{r+2}$, $y_{h}=d_{eff}\frac{1}{r+2}$.
The correlation length exponent follows also as $\nu=1/y_t$.
Note that the critical exponents calculated in the cases $0<r<2$ and $-1<r<0$ reach the same limit as $r\rightarrow 0$ ($\alpha\rightarrow -\infty$, $\beta\rightarrow\infty$, $\gamma\rightarrow 1$, $\delta\rightarrow 1$),
but are non-analytic at $r=0$, as demonstrated below.
The critical exponents for $-1<r<0$ are related to those for $0<r<2$ by replacing $r$ in the latter by $-r/(1+r)$, as seen in Fig. \ref{fig:exps} and Table \ref{tabler}.
The energy gap scales with $U^{1/|r|}$.

Surprisingly, the $r=0$ case requires separate considerations. The total energy acquires logarithmic corrections as\cite{gruner} 
\begin{gather}
\frac{E(\Delta)}{N}=-\left(\frac{1}{4W}-\frac{1}{U}\right)\Delta^{2}-\frac{\Delta^2}{2W}\mbox{ln}\left(\frac{2W}{\Delta}\right)
\end{gather}
close to the critical point, minimized by $\Delta=2We^{-\frac{2W}{U}}$, so $\beta=\infty$ and
$\frac{E_{min}(U)}{N}=-We^{-\frac{4W}{U}}$, hence $\alpha=-\infty$.
The correlation length scales as $1/\Delta$, implying $\nu=\infty$.
The exponential dependence of the order parameter and $\nu=\infty$ suggest that the conventional $r=0$ MF transition at $T=0$ as a function of the interaction and not temperature, resembles closely to a Kosterlitz-Thouless type 
of phase transition.
Indeed, both the Kondo temperature in the Kondo problem and the gap in the sine-Gordon model scales similarly, but not identically to the above findings.
The
magnetic field dependence of the order parameter reads as
\begin{equation}
s=\frac{h}{W}\mbox{ln}\left(\frac{h}{2W}\right),
\end{equation}
defining $\delta=1+0$, the "0" denoting the logarithmic corrections.
Interestingly, the linear response susceptibility diverges as $\sim 1/U^2$, giving $\gamma=2$.
This non-analytic behaviour is somehow expected, since all other exponents are non-analytic at the origin, although their limiting values at $r\rightarrow 0^\pm$ agree.

\begin{table}[h!]
\begin{tabular}{|c|c|c|c|c|c|}
\hline
exponents & $\alpha$ & $\beta$ & $\gamma$ & $\delta$ & $z$\\
\hline
$r\geq 2 $      &  0            & $\dfrac 12$      & 1 & 3                &  $\dfrac{d}{r+1}$ \\
$0<r<2$  & $1-\dfrac 2r$ & $\dfrac 1r$    & 1 & $r+1$            &  $\dfrac{d}{r+1}$ \\
$r=0$      &  $-\infty$    & $\infty$         & 2   & 1              & $d$\\
$-1<r<0$ & $3+\dfrac 2r$ & $-1-\dfrac 1r$ & 1 & $\dfrac{1}{r+1}$ &  $\dfrac{d}{r+1}$ \\
\hline
\end{tabular}
\caption{Summary of the MF quantum critical exponents for general gapless phases. The exponents for $-1<r<0$ can be obtained from the $0<r<2$ 
exponents by the $r\rightarrow -r/(r+1)$ replacement.
The $r>2$ exponents are identical to the conventional thermal MF exponents. Mono- and bilayer grahene corresponds to $r=1$ and 0 with $d=2$, respectively.}
\label{tabler}
\end{table}

In summary, we have studied the quantum critical properties of the MF transition in graphene. 
An ordered phase requires a critical coupling to occur, and the critical exponents listed in Table \ref{tab:exponents} differ from those in the Landau theory.
The former suggests that weak interaction cannot trigger a phase transition in graphene, and can only renormalize some 
of its parameters. This explains the lack of any experimentally verified phase transition so far.
Then, we have developed the MF theory of general bulk gapless phases, whose critical properties were found to vary strongly 
with the DOS exponent, summarized in Fig. \ref{fig:exps} and Table \ref{tabler}.
This extends the validity of the MF universality class into quantum criticality, covering a much  broader region than that of the conventional Landau theory.

This work was supported by the Hungarian Scientific Research Fund No. K72613 and by the Bolyai program of the
Hungarian Academy of Sciences.

\bibliographystyle{apsrev}
\bibliography{refgraph}
\end{document}